**Growth and Electrical Characterization of 2D Layered MoS$_2$/SiC Heterojunctions**

Edwin W. Lee II[1], Lu Ma[3], Digbijoy N. Nath[1], Choong Hee Lee, Yiying Wu[3], Siddharth Rajan[1,2]

[1]Department of Electrical and Computer Engineering, The Ohio State University, Columbus, Ohio 43210

[2]Department of Material Science and Engineering, The Ohio State University, Columbus, Ohio 43210

[3]Department of Chemistry and Biochemistry, The Ohio State University, Columbus, Ohio 43210

**Abstract**

The growth and electrical characterization of a heterojunction formed between 2D layered p-MoS$_2$ and nitrogen-doped 4H-SiC is reported. Direct growth of p-type MoS$_2$ films on SiC was demonstrated using chemical vapor deposition, and the MoS$_2$ films were found to be crystalline based on x-ray diffraction measurements. The resulting heterojunction was found to display rectification and current-voltage characteristics consistent with a p-n junction diode. Capacitance voltage measurements were used to determine the built-in voltage for the p-MoS$_2$/n-SiC heterojunction p-n diode. The demonstration of heterogeneous material integration between 2D layered semiconductors and 3D SiC enables a new class of heterostructures.

Wide band gap semiconductors are useful for a wide variety of applications including optoelectronics, high power, and high frequency. In the case of GaN, ZnO and SiC, p-doping has proven challenging. More recently, there has been a renewed interest in layered 2-dimensional materials such as graphene and metal dichalcogenides due to their interesting electronic properties, and their applications in transparent flexible electronics and highly scaled logic transistors. There has been particularly widespread interest in molybdenum disulfide ($MoS_2$) which has a non-zero band gap,[1] relatively high mobility (~200 $cm^2$/V-s),[2] and can be obtained from geological samples. $MoS_2$ has been proposed as a promising material for FETs,[3,4,5,6] sensors,[7,8] flexible electronics,[9] and other new devices.[10]

Integrating these layered transition metal dichalcogenides (TMDs) with wide band gap materials could open up applications with traditional 3D semiconductors that have yet to be achieved due to the lack of suitable p-type dopants. Figure 1 shows a lattice constant versus band gap chart for the TMDs and some of the wide band gap semiconductors.[11,12,13,14,15,16] The integration of 2D layered TMDs with III-nitrides could lead to very new and exciting device engineering flexibility. For example, integration of a low sheet resistance p-type layer on wide gap materials such as SiC or GaN could enable heterojunction bipolar transistors (HBTs) which have until now been challenging due to the low hole ionization and mobility in the wide band gap semiconductors. Such an idea was used to combine other materials such as GaAs and InGaAs with GaN to create wafer-fused devices,[17] with the objective of combining the low sheet resistance of the GaAs layer with the large band gap and breakdown field of GaN to achieve a device with higher power-frequency performance. However, out-of-plane bonding in traditional semiconductors makes the integration of mismatched materials challenging due to interfacial defects.

While this general class of heterostructures could have many applications, we have focused on p-n junctions since the hole ionization and vertical and lateral hole transport in MoS$_2$ are expected to be superior to the p-type species of many wide band gap materials. In this work we show that a high quality heterojunction can be formed between p-type MoS$_2$ and n-SiC. Using electrical measurements, we demonstrate an anisotype p-n junction formed between 2D layered materials and a 3D semiconductor.

MoS$_2$ was grown directly on nitrogen-doped 4H SiC substrates with 4° miscut (Cree) using a method previously reported for growth of MoS$_2$ on sapphire.[18] The commercially available SiC substrate was uniformly doped with n-type carrier concentration between 1-3 x 10$^{18}$ cm$^{-3}$. The substrates were degreased using ultrasonic dips in acetone, isopropanol and deionized water, then baked at 120°C for 5 minutes. Mo (25 Å)/Nb (2Å) /Mo (25Å) layers were deposited on substrates by sputtering (AJA Orion RF/DC Sputter Deposition Tool). 90 mg of MoS$_2$ powder was placed in a small quartz tube. The small quartz tube and the metalized substrate were inserted in a larger quartz tube (inner diameter of 1 cm). The open end of the large quartz tube was pumped down by a mechanical pump for 30 minutes, sealed and heated heated to 1100°C for 4.5 hours and then cooled down to room temperature at a rate of 0.5°/min.

The crystalline orientation and quality were evaluated using double-axis high-resolution X-ray diffraction measurements (Figure 2). The XRD spectra shows the characteristic (002), (006), and (008) peaks for MoS$_2$.[18] The crystallinity of MoS$_2$ on SiC suggests that, similar to previous work on sapphire, optimal growth conditions lead to high quality growth of layered transition metal dichalcogenides. Figure 3 shows a 5 μm x 5μm atomic force microscope (AFM) scan of p-MoS$_2$ on nitrogen-doped 4H SiC with 1.66 nm RMS roughness with a 12 nm height scale.

Standard lithography was used to pattern regions for ohmic contacts, and Ni/Au/Ni was deposited to form p-type ohmic contacts to the $MoS_2$. Isolation mesas were created using the previously deposited ohmic contacts as etch masks and $Ar/BCl_3$-based inductively coupled plasma/reactive ion etch (ICP/RIE) (30 W RIE power). Ohmic contact to n-type SiC was made by scratching the backside of the SiC substrate and applying an indium metal dot.

Current-voltage (I-V) and capacitance-voltage (C-V) measurements were performed using an Agilent B1500 semiconductor device analyzer equipped with medium power source/monitor units (MPSMUs) and multi-frequency capacitance measurement units (MPCMUs), for the respective measurements. Lateral transport measurements performed on unisolated p-$MoS_2$ between two Ni/Au/Ni contacts were linear as shown in Figure 4, suggesting ohmic conduction in the film. Transfer length measurements (TLM) gave a contact resistance of 3.03 Ω-mm (specific resistance $0.182 \times 10^{-5}$ Ω-mm$^2$) and sheet resistance of 2.07 kΩ/square. Hall measurements on similar Nb-doped samples grown by CVD on insulating sapphire substrates were found to give p-type carrier concentration of $3.1 \times 10^{20}$ cm$^{-3}$ and hole mobility of 8.5 cm$^2$/V-s.[19]

The vertical I-V characteristic was measured between an ohmic contact to p-$MoS_2$ (contact area 2500 μm$^2$) and indium metal dots pressed into the backside of the SiC substrate. Bias was applied to the p-$MoS_2$ ohmic contact with respect to the n-SiC contact. The I-V characteristic (Figure 5) showed approximately 7 orders of rectification (+/- 1.0 V) and a subthreshold slope of 108 mV/decade. The inset of figure 5 includes the measured device structure.

Capacitance-voltage measurements were performed and the double-sweep C-V characteristic (Figure 6a) showed negligible hysteresis, suggesting a low density of interface traps or defects. The C-V curve showed a typical characteristic as expected for a reverse biased pn-junction. Figure 6b shows $1/C^2$ as a function of applied bias corresponding to the device measured in Figure 6a. The linear curve shows that the doping on both sides was constant. The curve was extrapolated to determine its intercept on the voltage axis which corresponds with the built-in voltage of the junction. The $1/C^2$ fit suggests a built-in voltage of approximately 2.3 V for the p-$MoS_2$/n-SiC heterojunction.

In conclusion, we have demonstrated the ability to grow high quality, single crystal, p-doped $MoS_2$ on SiC by chemical vapor deposition. We characterized the heterojunction formed by $MoS_2$ and SiC with I-V and C-V measurements, showing that the $MoS_2$/SiC heterojunction is rectifying and that the built-in voltage is approximately 2.3 V. The integration of $MoS_2$ and SiC described in this work could lead to several promising heterostructure devices that integrate layered 2D and traditional 3D semiconductor materials.

**Acknowledgements:** E.L. and S.R. acknowledge NSF NSEC (CANPD) Program (EEC0914790) and NSF grant ECCS-0925529. L.M. and Y.W. acknowledge the support from NSF (CAREER, DMR-0955471). This work was supported in part by The Ohio State University Materials Research Seed Grant Program.

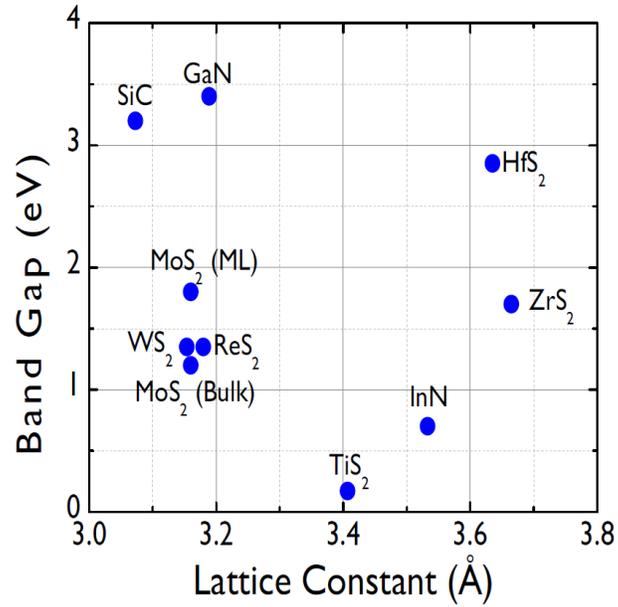

Figure 1: Lattice constant versus band gap diagram showing the transition metal dichalcogenides (TMDs) with several wide band gap semiconductors.

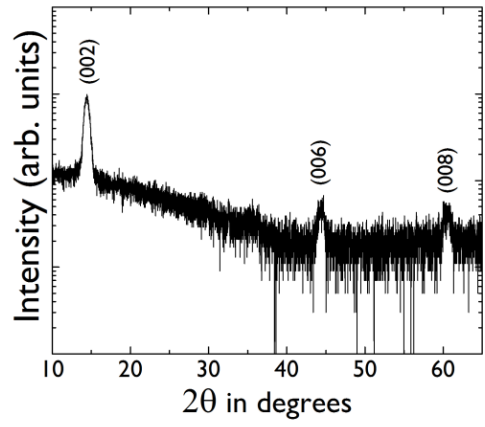

Figure 2: XRD scan of CVD-grown p-$MoS_2$ on SiC including the characteristic (002), (006), and (008) peaks of $MoS_2$.

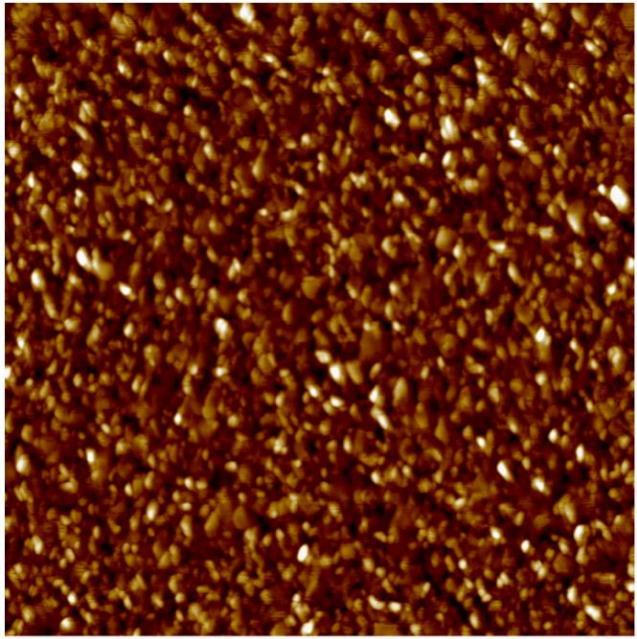

Figure 3: 5 μm x 5 μm scan AFM of p-MoS$_2$ on nitrogen-doped 4H SiC with 1.66 nm RMS roughness and a height scale of 12 nm.

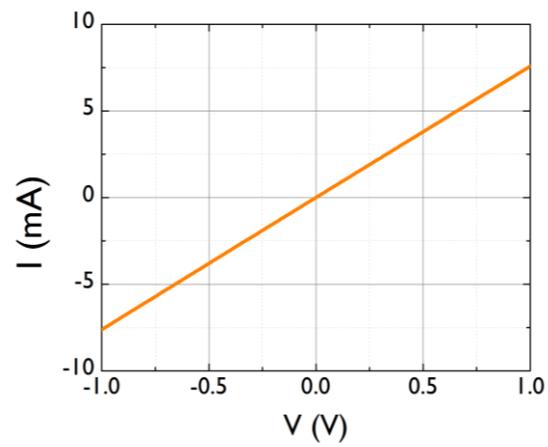

Figure 4: Lateral I-V measurements displaying ohmic contact to MoS$_2$.

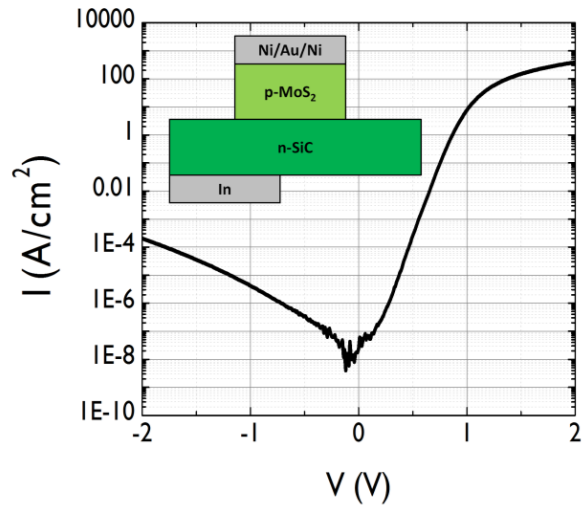

Figure 5: Vertical I-V is measured with bias applied to the p-MoS$_2$ contact. The inset shows the measured device structure.

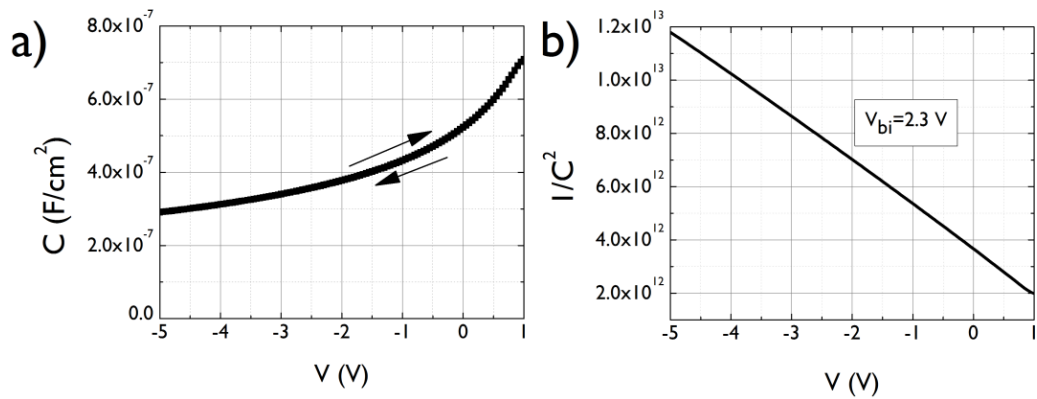

Figure 6: a) Capacitance voltage characteristic of the p-MoS$_2$/n-SiC heterojunction. The lack of hysteresis in the double-sweep C-V is indicative of a high-quality heterointerface. b) $1/C^2$ plotted with respect to applied bias. The voltage axis intercept corresponds with a built-in voltage of approximately 2.3 V.